\documentclass[aps,prb,showkeys,noshowpacs,superscriptaddress,twocolumn,floatfix]{revtex4}
\usepackage{graphicx}
\usepackage{amsmath}
\usepackage{amsfonts}

\begin{document}

\title{Truth seekers in opinion dynamics models}

\author{Krzysztof Malarz}
\homepage{http://home.agh.edu.pl/malarz/}
\email{malarz@agh.edu.pl}
\affiliation{Faculty of Physics and Applied Computer Science,
AGH University of Science and Technology,\\
al. Mickiewicza 30, PL-30059 Krak\'ow, Euroland}
\affiliation{Institute of Theoretical Physics,
University of Cologne, D-50923 K\"oln, Euroland}

\begin{abstract}
We modify the model of Deffuant {\em et al.} to distinguish true opinion among others in the fashion of Hegselmann and Krause ({\tt http://jasss.soc.surrey.ac.uk/9/3/10.html}).
The basic features of both models modified to account for truth seekers are qualitatively the same.
\end{abstract}



\maketitle

{\bf Introduction.} Opinion dynamics simulations \cite{opinion,sznajd,hegselmann,deffuant} (see Ref. \onlinecite{review} for review) seem to be the most fashionable part of sociophysics \cite{book}.
When your girlfriend asks you `Do I look fine?' you may use the Sznajd model \cite{sznajd} to answer this general question as Sznajd agents behave similarly to magnetic spins: they are allowed to have only two opinions $s_i\in\{0,1\}$ --- for example `yes' or `no'.
But when the question is open --- for instance: `How am I looking like today?' --- the space of possible answers enlarges drastically.
The correct answers are `Super!' and `Exceptionally!' but not `O.K.' or even `nice'.
The latter is reserved for the same question but about a pretty girl starring in any Hollywood movie.
For modeling such opinions the Hegselmann--Krause \cite{hegselmann} (H-K) or Deffuant {\em et al.} \cite{deffuant} models are more appropriate as they offer a continuous interval of possible opinions $s_i\in[0,1]$.

Now, you and your interlocutor may have a wrong opinion about your girlfriend.
The truth may be elsewhere.
The modifications of H-K model which allow to introduce truth seekers among exchanging their opinions agents were presented very recently in Ref. \onlinecite{hegselmann2006}.

In this paper we would like to check if the same is available for Deffuant {\em et al.} model \cite{deffuant} were Assmann \cite{assmann} had already introduced a multitude of truths.

{\bf The model.} In the original Deffuant {\em et al.} model two persons (let say $i$ and $j$) exchange their opinion about given topic if their current opinions do not differ more than {\em confidence level} $\varepsilon$, i.e. when $|s_i-s_j|\le\varepsilon$.
In such case, after discussion their change their opinions slightly, i.e.
\begin{equation}
\begin{cases}
s_i \to & s_i +\mu \delta \\
s_j \to & s_j -\mu \delta 
\end{cases},
\label{def}
\end{equation}
where $\delta=s_j-s_i$ and $\mu\in[0,1/2]$ describes a speed of opinion changes.
If their opinions are too distant, i.e. $|\delta| > \varepsilon$, the agents do not change their opinions at all.

To account for the true opinion Hegselmann and Krause introduced two additional parameters: $T\in [0,1]$ and $\alpha_i$ --- which represent the {\em true} opinion and the {\em strength of the attraction to the truth} for $i$-th agent, respectively \cite{hegselmann2006}.

With these two additional terms Eq. \eqref{def} for the Deffuant {\em et al.} model becomes
\begin{equation}
\begin{cases}
s_i \to & s_i + \mu [ \alpha_i(T-s_i) +(1-\alpha_i) \delta ] \\
s_j \to & s_j + \mu [ \alpha_j(T-s_j) -(1-\alpha_j) \delta ]
\end{cases}.
\label{mod}
\end{equation}

The case $\alpha_i=0$ $(i=1,\cdots,N)$ corresponds to the original Deffuant {\em et al.} model and for $\alpha_i=1$ $(i=1,\cdots,N)$ agents do not exchange opinions each to other but tends towards the true one.

\begin{figure*}
\begin{center}
\includegraphics[width=0.45\textwidth]{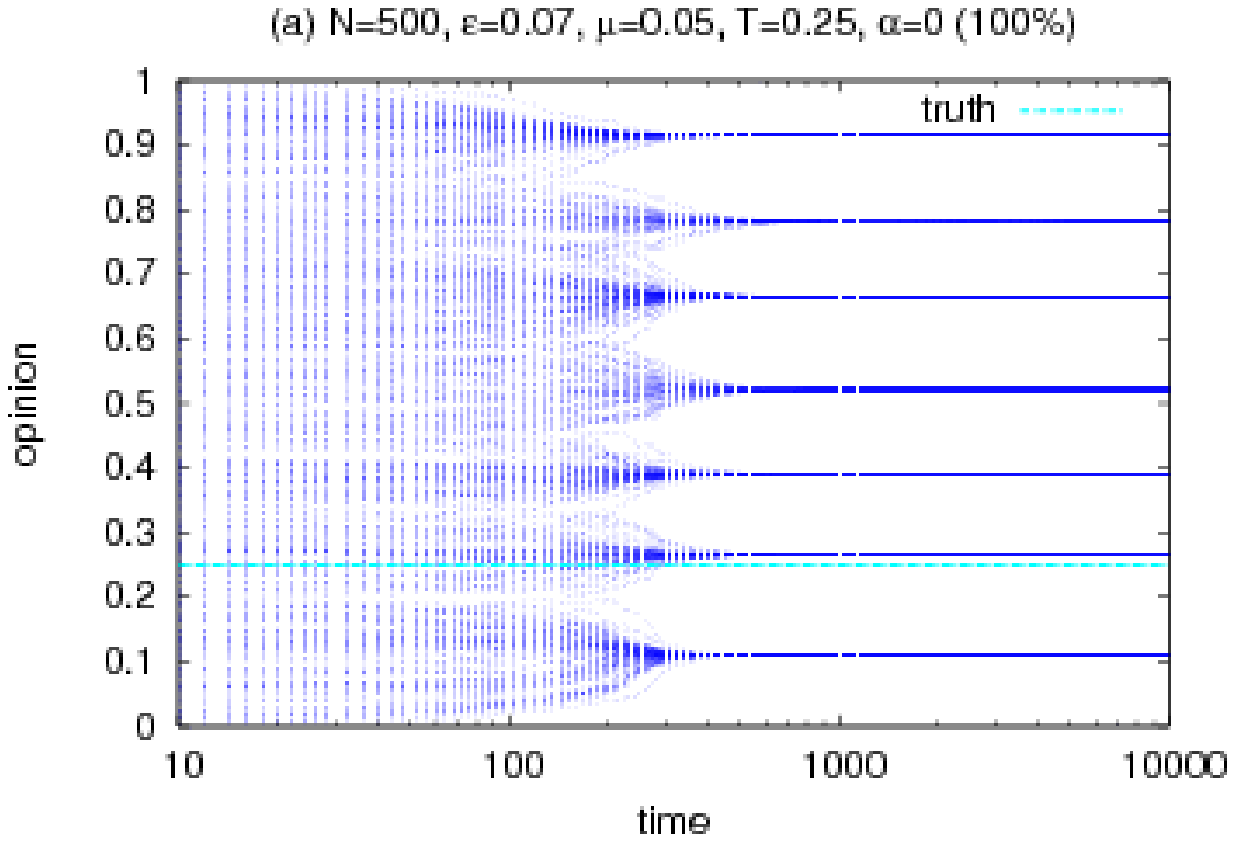}
\includegraphics[width=0.45\textwidth]{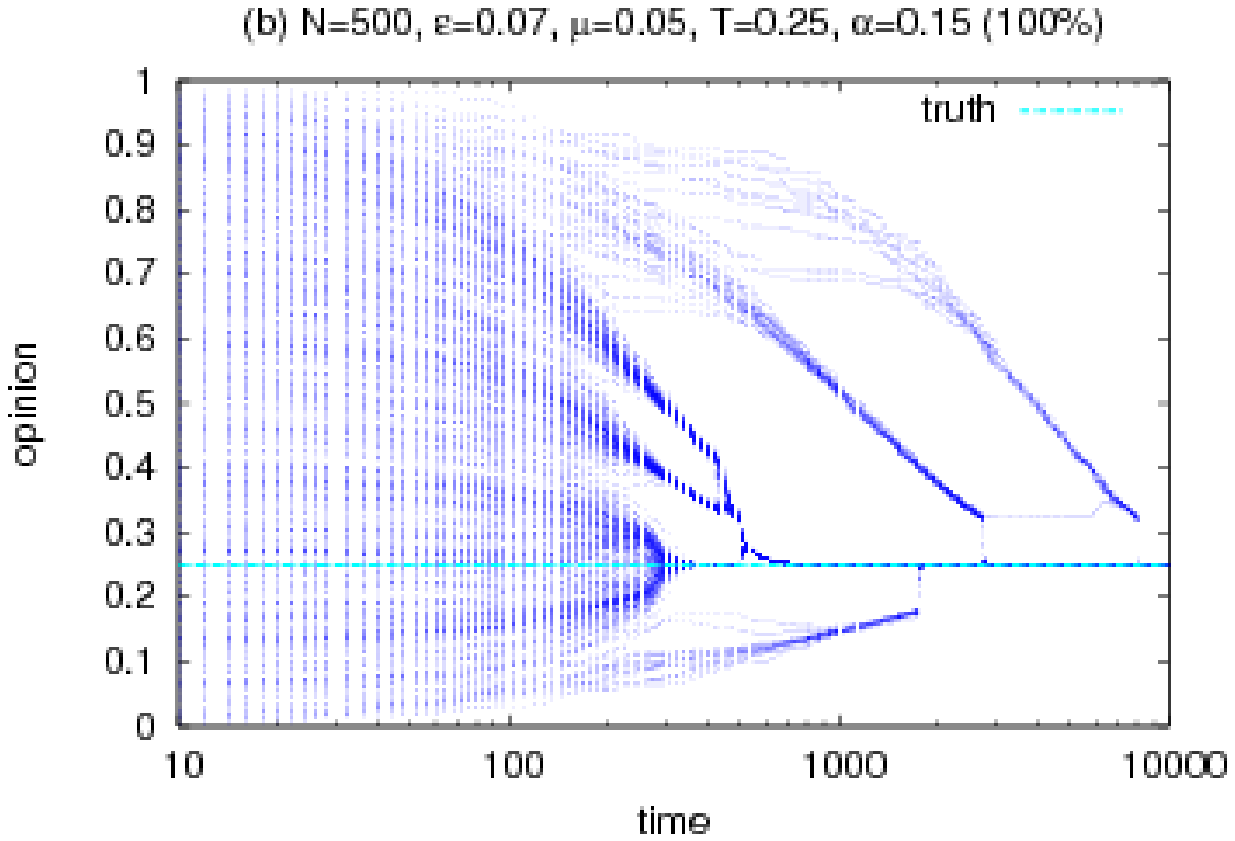}\\
\includegraphics[width=0.45\textwidth]{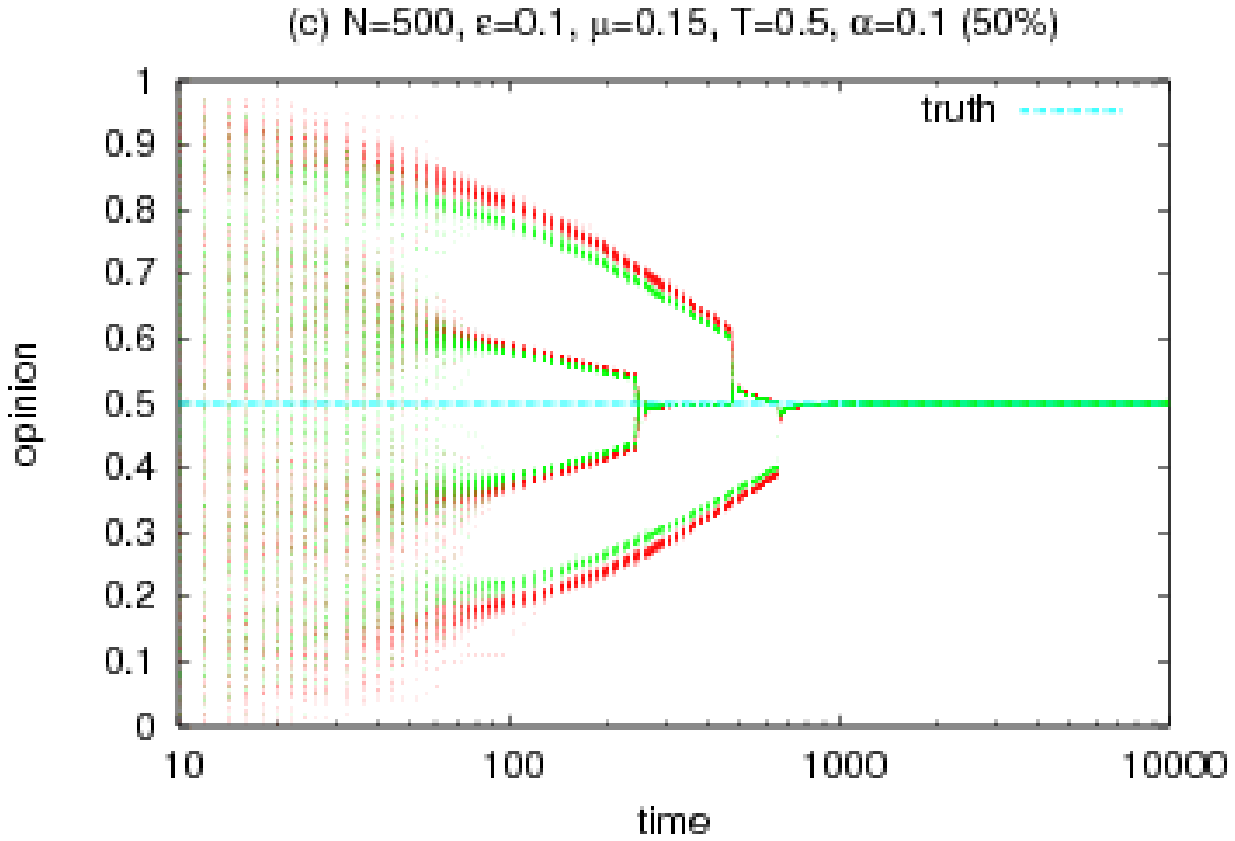}
\includegraphics[width=0.45\textwidth]{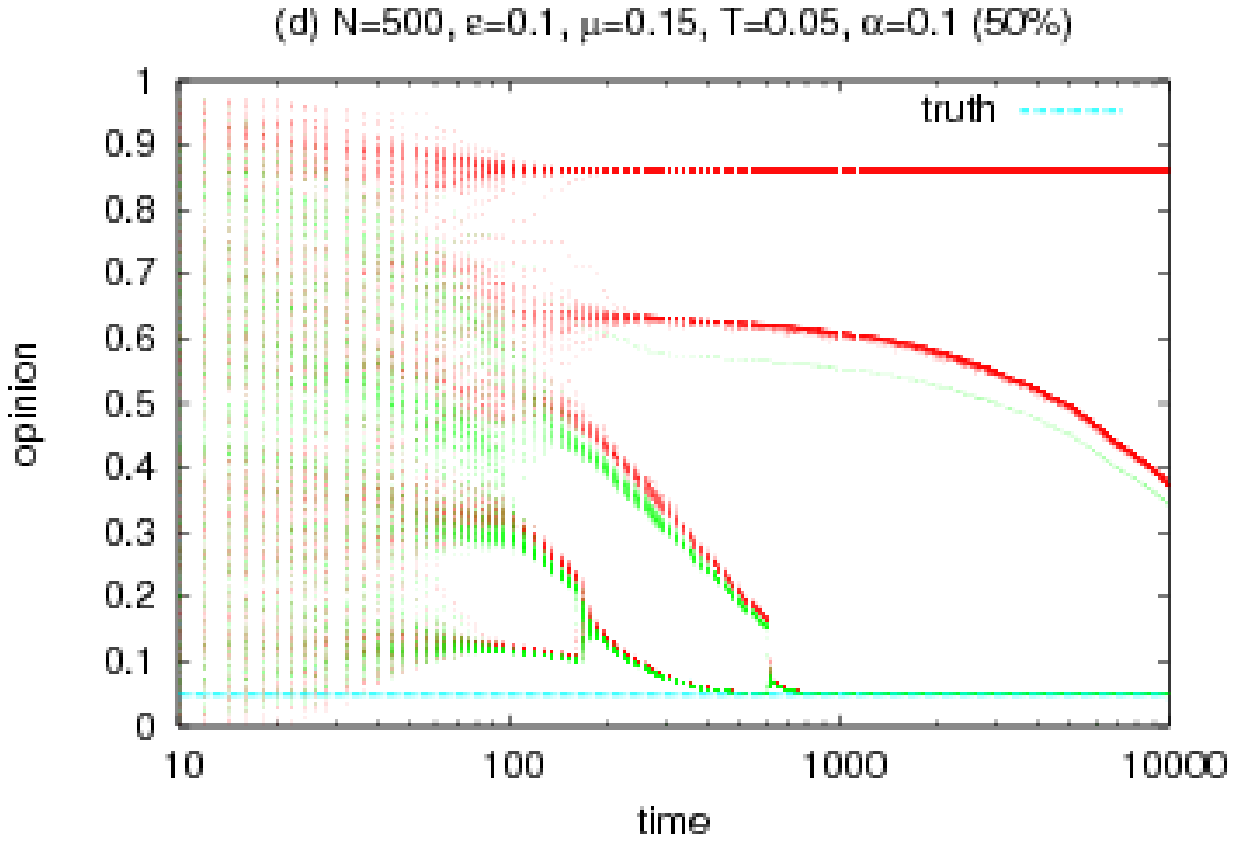}\\
\includegraphics[width=0.45\textwidth]{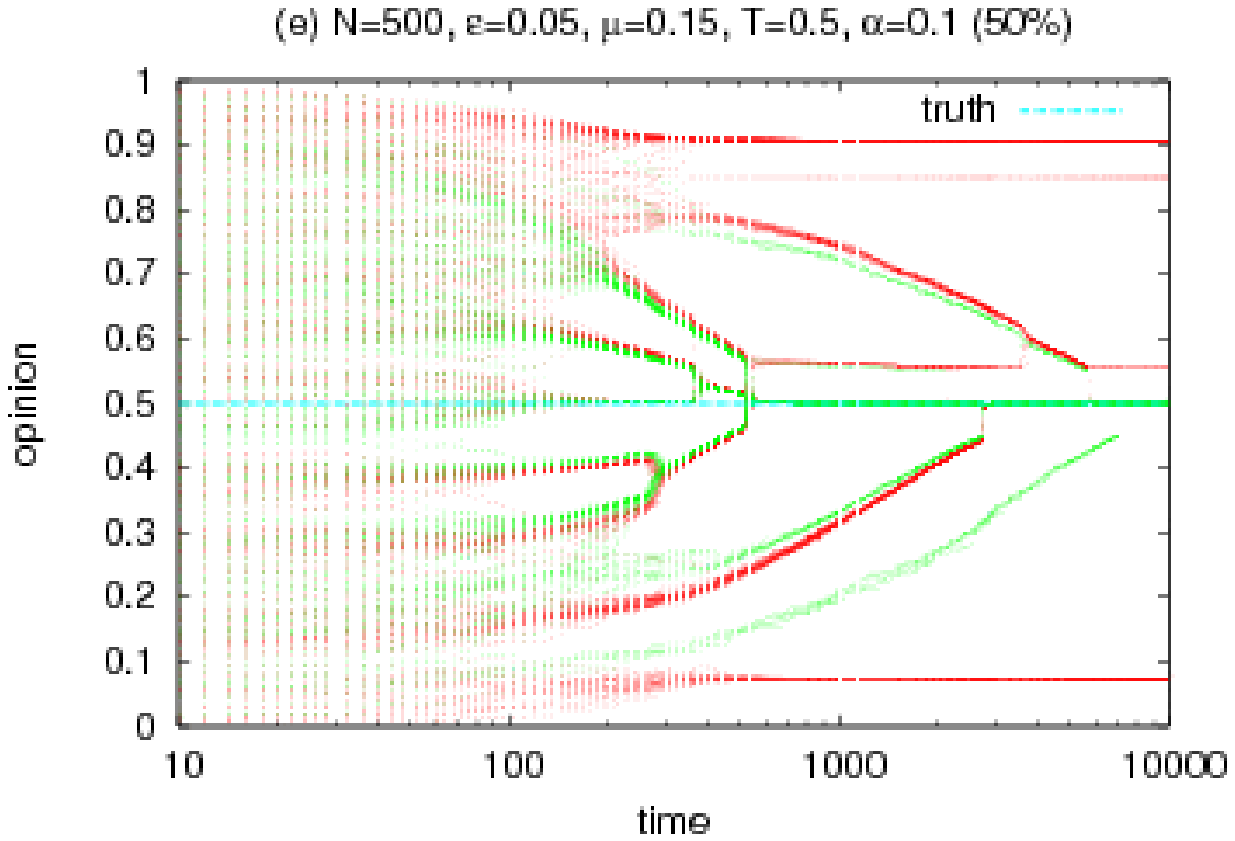}
\includegraphics[width=0.45\textwidth]{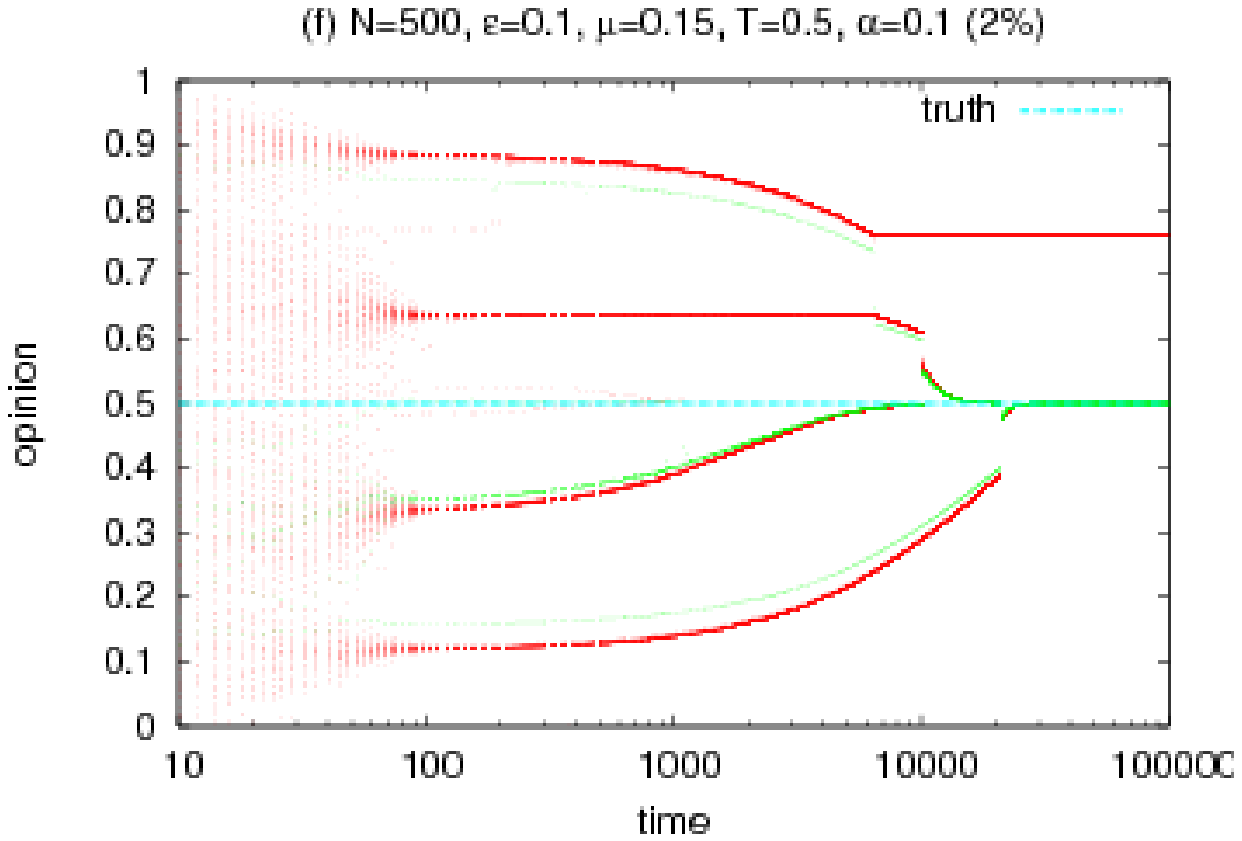}
\end{center}
\caption{(Color online). A few examples of opinion dynamics for Deffuant model with truth seekers. 
The model parameters and given in sub-figures captions.
The straight horizontal line indicates the `truth'.}
\label{fig}
\end{figure*}

{\bf The results.} In Fig. \ref{fig} the results of simulation are presented for opinion dynamics of $N=500$ agents with initially randomly chosen opinions $s_i$ (the same for all sub-figures).
The model parameters are shown in the headline of all sub-figures.
In Fig. \ref{fig}(a) the opinion dynamics governed by original Deffuant {\em et al.} model is presented ($\alpha_i=0$ for all $i$). 
In Fig. \ref{fig}(b) all agents are the truth seekers.
In Figs. \ref{fig}(c)--(e) half of agents is the truth seeker ($\alpha_i>0$, marked as green) while the second half is not ($\alpha_i=0$, marked as red).
In Fig. \ref{fig}(f) only 2\% agents search for the truth.
Parts c/d, c/e, c/f differ {\em only} with $T$, $\varepsilon$ and the fraction of the truth seekers, respectively.
For $N=10^2$ and $10^3$ the results are qualitatively the same.

The obtained results support the observation from Figs. 1--8 in Ref. \onlinecite{hegselmann2006}.

In conclusions, the Deffuant {\em et al.} model \cite{deffuant} with necessary modifications which allow to simulate true opinion among others gives qualitatively the same results as the H-K model modified for the same purpose \cite{hegselmann2006}.

{\bf Acknowledgments.} Author is grateful to Dietrich Stauffer for his hospitality in K\"oln and to EU grant GIACS. 
Part of calculation was carried out in ACK-CYFRONET-AGH.
Time on HP Integrity Superdome is financed with grant no. MNiI/HP\_I\_SD/AGH/047/2004.


\end{document}